\documentclass[prc,twocolumn,superscriptaddress,showpacs,twoside,floatfix]
{revtex4-1}

\usepackage{amssymb,epsfig}

\begin{document}

\title{$^{10}$He low-lying states structure uncovered by correlations}

\author{S.I.~Sidorchuk}
\affiliation{Flerov Laboratory of Nuclear Reactions, JINR, Dubna, RU-141980
Russia}
\author{A.A.~Bezbakh}
\affiliation{Flerov Laboratory of Nuclear Reactions, JINR, Dubna, RU-141980
Russia}
\author{V.~Chudoba}
\affiliation{Flerov Laboratory of Nuclear Reactions, JINR, Dubna, RU-141980
Russia}
\address{Institute of Physics, Silesian University in Opava, Bezru\v{c}ovo
n\'{a}m.\ 13, 74601 Czech Republic}
\author{I.A.~Egorova}
\address{Bogolyubov Laboratory of Theoretical Physics, JINR, Dubna, 141980
Russia}
\author{A.S.~Fomichev}
\affiliation{Flerov Laboratory of Nuclear Reactions, JINR, Dubna, RU-141980
Russia}
\author{M.S.~Golovkov}
\affiliation{Flerov Laboratory of Nuclear Reactions, JINR, Dubna, RU-141980
Russia}
\author{A.V.~Gorshkov}
\affiliation{Flerov Laboratory of Nuclear Reactions, JINR, Dubna, RU-141980
Russia}
\author{V.A.~Gorshkov}
\affiliation{Flerov Laboratory of Nuclear Reactions, JINR, Dubna, RU-141980
Russia}
\author{L.V.~Grigorenko}
\affiliation{Flerov Laboratory of Nuclear Reactions, JINR, Dubna, RU-141980
Russia}
\affiliation{GSI Helmholtzzentrum f\"{u}r Schwerionenforschung, Planckstra{\ss}e
1, D-64291 Darmstadt, Germany}
\affiliation{National Research Centre ``Kurchatov Institute'', Kurchatov sq.\
1, RU-123182 Moscow, Russia}
\author{G.~Kaminski}
\affiliation{Flerov Laboratory of Nuclear Reactions, JINR, Dubna, RU-141980
Russia}
\address{Institute of Nuclear Physics PAN, Radzikowskiego 152, PL-31342
Krak\'{o}w, Poland}
\author{S.A.~Krupko}
\affiliation{Flerov Laboratory of Nuclear Reactions, JINR, Dubna, RU-141980
Russia}
\author{E.A.~Kuzmin}
\affiliation{National Research Centre ``Kurchatov Institute'', Kurchatov sq.\
1, RU-123182 Moscow, Russia}
\author{E.Yu.~Nikolskii}
\affiliation{National Research Centre ``Kurchatov Institute'', Kurchatov sq.\
1, RU-123182 Moscow, Russia}
\affiliation{RIKEN Nishina Center, Hirosawa 2-1, Wako, Saitama 351-0198, Japan}
\author{Yu.Ts.~Oganessian}
\affiliation{Flerov Laboratory of Nuclear Reactions, JINR, Dubna, RU-141980
Russia}
\author{Yu.L.~Parfenova}
\affiliation{Flerov Laboratory of Nuclear Reactions, JINR, Dubna, RU-141980
Russia}
\affiliation{Skobel'tsyn Institute of Nuclear Physics, Moscow State University,
119991 Moscow, Russia}
\author{P.G.~Sharov}
\affiliation{Flerov Laboratory of Nuclear Reactions, JINR, Dubna, RU-141980
Russia}
\author{R.S.~Slepnev}
\affiliation{Flerov Laboratory of Nuclear Reactions, JINR, Dubna, RU-141980
Russia}
\author{S.V.~Stepantsov}
\affiliation{Flerov Laboratory of Nuclear Reactions, JINR, Dubna, RU-141980
Russia}
\author{G.M.~Ter-Akopian}
\affiliation{Flerov Laboratory of Nuclear Reactions, JINR, Dubna, RU-141980
Russia}
\author{R.~Wolski}
\affiliation{Flerov Laboratory of Nuclear Reactions, JINR, Dubna, RU-141980
Russia}
\address{Institute of Nuclear Physics PAN, Radzikowskiego 152, PL-31342
Krak\'{o}w, Poland}
\author{A.A.~Yukhimchuk}
\affiliation{All-Russian Research Institute of Experimental Physics, RU-607190 Sarov,
Russia}
\author{S.V.~Filchagin}
\affiliation{All-Russian Research Institute of Experimental Physics, RU-607190 Sarov,
Russia}
\author{A.A.~Kirdyashkin}
\affiliation{All-Russian Research Institute of Experimental Physics, RU-607190 Sarov,
Russia}
\author{I.P.~Maksimkin}
\affiliation{All-Russian Research Institute of Experimental Physics, RU-607190 Sarov,
Russia}
\author{O.P.~Vikhlyantsev}
\affiliation{All-Russian Research Institute of Experimental Physics, RU-607190 Sarov,
Russia}

\date{\today. {\tt File: he10-10.tex}}

\begin{abstract}
The $0^+$ ground state of the $^{10}$He nucleus produced in the $^3$H($^8$He,$p$)$^{10}$He
reaction was found at about $2.1\pm0.2$ MeV ($\Gamma \sim 2$ MeV) above the three-body
$^{8}$He+$n$+$n$ breakup threshold.  Angular correlations observed for $^{10}$He
decay products show prominent interference patterns allowing to draw conclusions
about the structure of low-energy excited states. We interpret the observed correlations
as a coherent superposition of the broad $1^-$ state having a maximum at energy $4-6$ MeV
and the $2^+$ state above 6 MeV, setting both on top of the $0^+$ state ``tail''.
This anomalous level ordering indicates that the breakdown of the $N=8$ shell known in
$^{12}$Be thus extends also to the $^{10}$He system.
\end{abstract}


\maketitle

%
%===============================================================================
%
\textit{Introduction.}
%
%===============================================================================
%
With the improvement of knowledge about the nuclei far from the ``stability valley''
and with the development of experimental techniques the interests of researchers are naturally
shifting to the regions beyond the nuclear stability lines. The understanding of
such systems is indispensable for deeper insights into the nuclear dynamics, for further development
of nuclear models, and for nuclear astrophysics applications. Among the isotopes observed as
resonances $^{10}$He has the largest $N/Z$ ratio on nuclear chart, thus
representing the most extreme nuclear matter asymmetry. According to natural shell-model
considerations $^{10}$He should be the second lightest double-magic nucleus after $^{4}$He.
However, the search for nuclear-stable $^{10}$He was in vain and in 1994 it was observed as a
resonance \cite{Korsheninnikov:1994}. Thus an additional stabilizing effect of shell closure
was not observed. Our new results cast even more doubt in magic nature of this nucleus giving less
binding than expected \cite{Korsheninnikov:1994,Johansson:2010,Johansson:2010b} and providing
evidence for the low-lying negative parity intruder state.

This isotope is difficult to study as there are very few ways to produce the nucleus with such
an enormous neutron excess. There are several qualitatively different experimental results on the
$^{10}$He spectrum. A \textit{broad} ($\Gamma \lesssim 1.2$ MeV) resonance at energy $E_T=1.2(3)$ MeV
was observed \cite{Korsheninnikov:1994} as a result of proton knockout from $^{11}$Li ($E_T$ is the energy
relative to the three-body $^8$He+$n$+$n$ decay threshold). A very similar excitation spectrum was
obtained for $^{10}$He in the analogous reaction at higher beam energy \cite{Johansson:2010}.
However, the authors of this work came to somewhat different parameters for the $0^+$ $^{10}$He ground
state (g.s.) resonance: \ $E_T=1.5$ MeV and $\Gamma=1.9$ MeV. Also the existence of $2^+$
excited state was inferred with $E_T=4$ MeV and $\Gamma=1.6$ MeV basing on three-body $^8$He+$n$+$n$
correlations \cite{Johansson:2010b}. The authors of Ref. \cite{Ostrowski:1994} reported a \textit{narrow}
($\Gamma=0.3$ MeV) $^{10}$He $0^+$ g.s.\ at $E_T=1.07(7)$ MeV populated in a double charge-exchange reaction.
Besides, the two peaks at 4.3 and 7.9 MeV \cite{Ostrowski:1994} were interpreted as excited
states of $^{10}$He with spin-parities $2^+$ and $3^-$ .

Neutron transfer is known to be a reliable tool for the study of nuclear systems with large
neutron excess. No $^{10}$He resonance peak in the vicinity of $E_T\sim1$ MeV was found
in the 2n transfer reaction $^3$H($^8$He,$p$)$^{10}$He \cite{Golovkov:2009}. The observed
$^{10}$He spectrum showed a \textit{broad} group of events at $2.5-4.5$ MeV. In this Letter
we report about a refined experiment on the $^3$H($^8$He,$p$)$^{10}$He reaction which
gives the $^{10}$He g.s.\ position at $E_T \sim 2.1$ MeV. The new work comprises a convincing
statistics with correlation results enabling spin-parity assignments.

%-------------------------------------------------------------------------------
\begin{figure}[t]
\centerline{\includegraphics[width=0.45\textwidth]{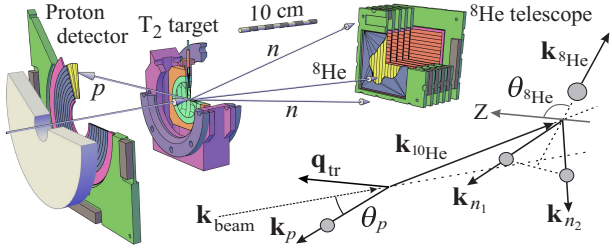}}
\caption{Experimental layout and kinematical scheme for the $^3$H($^{8}$He,$p$)$^{10}$He reaction.
The proton and $^{10}$He momenta are shown in the lab system, the variables applied to the $^{10}$He
decay products are presented in the $^{10}$He center-of-mass system with Z axis parallel to the
transferred momentum vector.}
\label{fig:layout}
\end{figure}
%-------------------------------------------------------------------------------

%===============================================================================
%
\textit{Experiment.}
%
%===============================================================================
%
A primary $^{11}$B beam with energy 36A MeV was delivered by the U-400M cyclotron (JINR, Dubna).
The secondary 21.5A MeV beam of $^8$He with intensity $\sim1.5\cdot10^{4}$ $s^{-1}$ obtained with
the fragment-separator ACCULINNA \cite{Rodin:1997} hit a gaseous tritium target \cite{Yukhimchuk:2003},
see Fig.\ \ref{fig:layout}. The two thin plastic scintillators set on a 8 m base before the target
allowed the beam particle identification and the time-of-flight (TOF) measurement with an accuracy
of about 0.5 ns. The two multiwire proportional chambers performed tracking for incoming $^8$He ions
providing hit positions on the target cell with accuracy $\sim$1.5 mm. The target windows of 25 mm
in diameter were sealed with two pairs of 8.4 $\mu$ stainless steel foils. The 6 mm thick target cell was
filled with tritium with a gas pressure of 0.92 atm and cooled down to 26 K. The total integral
flux of $^8$He was $\sim1.4\cdot10^{10}$. The concept of the experiment was similar to that applied
in our previous works \cite{Golovkov:2004,Golovkov:2005,Golovkov:2007}. This approach implies the
detection of recoil protons emitted from the target in backward direction in the lab system.
This low-background kinematical range corresponds to small angles in the center-of-mass (c.m.)
system. The protons were detected with a 1 mm thick annular Si detector with the inner and
outer diameters of its sensitive area of 32 and 82 mm, respectively. The two detector
sides were segmented in 16 rings and 16 sectors. With the proton detector installed 100 mm upstream
the target the $^{10}$He missing mass was measured with resolution of about 0.5 MeV (FWHM). This estimate
followed from a Monte-Carlo simulation and was found to be in a good agreement with the results obtained
for the $^3$H($^{6}$He,$p$)$^{8}$He reaction populating the well known 0$^+$ and 2$^+$ states of $^8$He.
A telescope composed of six square, 1 mm thick, $61\times61$ mm$^2$ Si detectors (having 16 strips each)
was placed 250 mm downstream the target to detect the $^8$He fragments originating from the $^{10}$He decay
in coincidence with the recoil protons.

\begin{figure}[t]
\centerline{\includegraphics[width=0.38\textwidth]{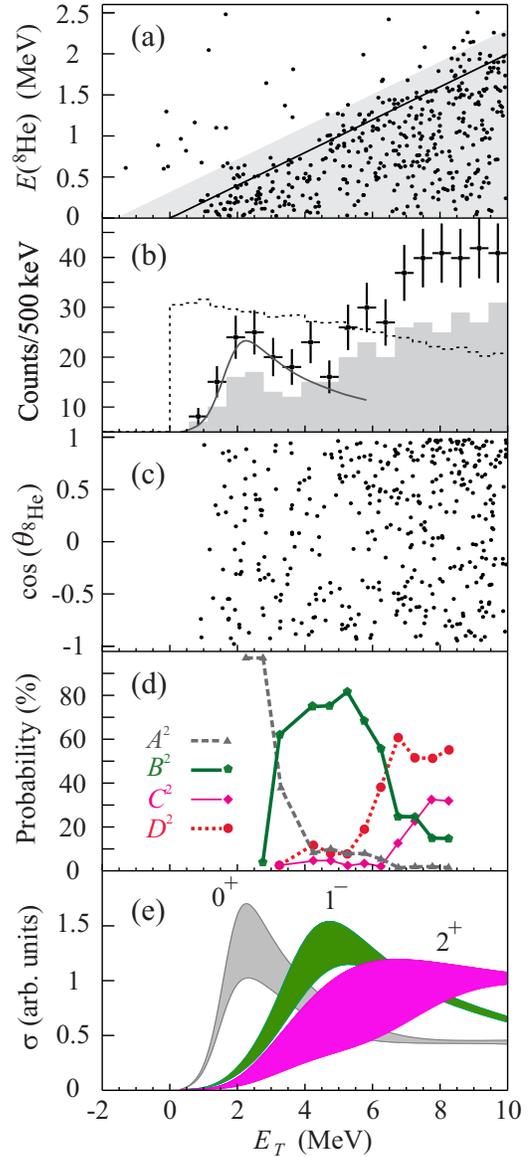}}
\caption{(a) Scatter plot of $E$($^8$He) vs.\ $E_T$. (b) $^{10}$He missing mass spectrum.
Points with error bars correspond to the total bulk of events, while the grey histogram
is obtained under condition $\varepsilon<0.5$. The dotted histogram shows the detection
efficiency. The theoretical curve from the panel (e) is given to guide eye.
(c) Angular distribution of $^{8}$He in the $^{10}$He c.m.\ frame. (d) Squared amplitudes
(they do not change signs) of different partial contributions in Eq.\ (\ref{eq:polinomial})
deduced from the angular distribution. (e) Theoretically predicted $^{10}$He spectra for
different $J^{\pi}$. Shaded areas reflect the uncertainty of these
calculations.}
\label{fig:results}
\end{figure}

%===============================================================================
%
\textit{``Triangle'' presentation of the data.}
%
%===============================================================================
%
The energy $E$($^8$He) of $^{8}$He in the c.m.\ of $^{10}$He is plotted vs.\ $E_T$
in Fig.\ \ref{fig:results} (a). This presentation allows to estimate background conditions
and reject events located outside the kinematically allowed region. Events appearing below the
solid line in Fig.\ \ref{fig:results} (a) satisfy the condition $E(^{8}\text{He})<E_T/5$. To take
into account the experimental resolution we present below results obtained for events
located inside the broader shaded triangle. The background inside the triangle was measured
in irradiations made with empty target and was found to be negligible.

%-------------------------------------------------------------------------------
%-------------------------------------------------------------------------------

%===============================================================================
%
\textit{Missing mass of $^{10}$He.}
%
%===============================================================================
%
The projected missing mass spectrum from the data of  Fig.\ \ref{fig:results} (a) is shown by points
with error bars in  Fig.\ \ref{fig:results} (b). The $^{10}$He g.s.\ peak is clearly seen at
about 2.1 MeV. Above the g.s.\ the spectrum is quite featureless showing a smooth rise after 4 MeV.
Superimposed on the experimental points in Fig.\ \ref{fig:results} (b) is the
$J^\pi=0^+$ g.s. spectrum of $^{10}$He theoretically predicted in Ref. \cite{Grigorenko:2008}.
Note the good correspondence between experimentally observed spectrum and theory within the
main part of the $^{10}$He g.s.\ peak.

%-------------------------------------------------------------------------------
\begin{figure}[t]
\centerline{\includegraphics[width=0.4\textwidth]{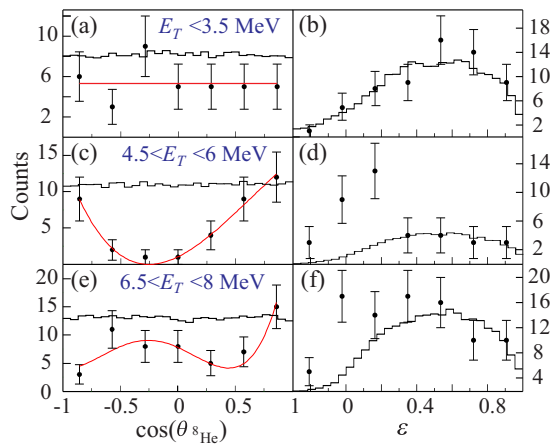}}
\caption{Panels (a,c,e) show the angular distributions of $^{8}$He measured in
different energy ranges of $^{10}$He excitation. Curves show fits made by using
Eq.\ (\ref{eq:polinomial}). Panels (b,d,f) present energy distributions between
neutrons obtained for the same energy ranges. Dots with error bars are experimental
data, the histograms show the detection system response to the phase volume.}
\label{fig:angdistr}
\end{figure}
%-------------------------------------------------------------------------------

%===============================================================================
%
\textit{Angular distribution of $^{8}$He.}
%
%===============================================================================
%
Angular correlations obtained for $^8$He emitted from $^{10}$He were analyzed using a specific
frame with Z axis coinciding in the direction with the transferred momentum vector
$\mathbf{q}_{\text{tr}}=\-(1/4)\mathbf{k}_{\text{beam}}-\mathbf{k}_{\text{p}}$.
The angular distribution of $^{8}$He vs.\ $^{10}$He decay energy is shown in Fig.\ \ref{fig:results} (c).
Three regions with prominent and qualitatively different correlation patterns are seen in this plot:
(i) ``$s$-wave'' range $E_T<4$ MeV,  (ii) ``$s/p$ interference'' range $4<E_T<6$ MeV, (iii) ``$s/p/d$
interference'' range $6<E_T<8$ MeV. The angular distributions for these energy ranges obtained under
the condition $\varepsilon=E_{nn}/E_T<0.5$ ($E_{nn}$ is the $n-n$ relative energy) are shown in
Fig.\ \ref{fig:angdistr} (a,c,e) together with fits obtained by the expression
\begin{equation}
w=\left[AP_0(x)+B\sqrt{3}P_1(x)+C\sqrt{5}P_2(x)\right]^2+D^2.
\label{eq:polinomial}
\end{equation}
Here $P_l$ are Legendre polynomials with $x=\cos(\theta_{^8\text{He}})$.
Coefficients $A$, $B$ and $C$ are the amplitudes of coherent $s$-, $p$- and $d$-wave
contributions, respectively, while $D$ takes into account a decoherent ``background''.
The energy behavior of these amplitudes is presented in Fig.\ \ref{fig:results} (d).
We put an additional condition $\varepsilon <0.5$ as in the limiting case  $\varepsilon
\rightarrow 1$ the angle $\theta_{^8\text{He}}$ becomes degenerate. It is also
obvious that at $\varepsilon$ close to unity this angle is poorly defined from
data due to errors in the momentum reconstruction.

Note the region (ii) [Fig.\ \ref{fig:angdistr} (c)] where the distribution
tends to zero around small $|\cos(\theta_{^8\text{He}})|$ indicating that only
\textit{coherent contributions} take place in this energy range. Why there are such
expressed correlation patterns for $^{8}$He fragment distribution and how they could be
connected to the quantum numbers of the whole $^{10}$He system? In our analysis we base
on earlier experience obtained in analogous correlation studies of the three-body decay of
the $^{5}$H system ($t$+$n$+$n$ channel) populated in the ($t$,$p$) transfer reaction
\cite{Golovkov:2004,Golovkov:2005}. In such three-body systems prominent
correlation patterns could be formed if the reaction mechanism is one-step
(a direct reaction mechanism) and the transferred spin is zero ($\Delta S=0$).
For such conditions the formed correlations could be revealed in the frame where
Z axis coincides with the transferred momentum vector, because only zero magnetic
substates of orbital momentum are transferred resulting in the population of completely
aligned configurations in the final state.

%-------------------------------------------------------------------------------
\begin{figure}[t]
\centerline{\includegraphics[width=0.45\textwidth]{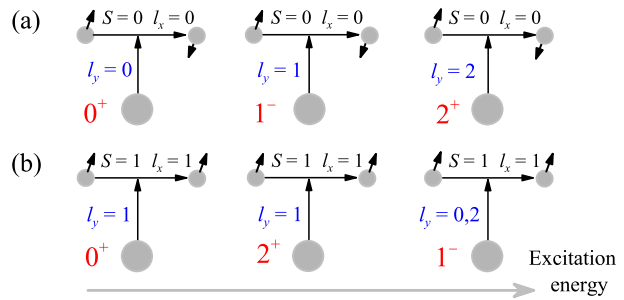}}
\caption{Expected major components of $^{10}$He wave function with different $J^{\pi}$
in the cluster $^{8}$He+$n$+$n$ representation. The rows (a) and (b) demonstrate the
two sets of components which may interfere in the angular distributions of $^{8}$He in the
$^{10}$He c.m.\ frame. States with different $J^{\pi}$ are ordered by energy
position expected for a particular configuration.}
\label{fig:tsys}
\end{figure}
%-------------------------------------------------------------------------------

To understand how the alignment of the whole three-body system is converted
into the expressed correlation patterns for the selected $^{8}$He fragment one needs to consider
the structure of the $^{10}$He states. There are two sets of possible major
configurations for the $^{10}$He wave function with different $J^{\pi}$ which interfere
with each other, see Fig.\ \ref{fig:tsys}. The set with $S=1$ can be rejected for the following reasons:
(i) the transfer of two neutrons with $S=1$ configuration is very unlikely. Extensive experience gained in
($t$,$p$) reaction studies points to strong dominance of ``dineutron'' ($S=0$ ``particle'') transfer;
(ii) $^{8}$He in $p$-wave configuration ($l_y=1$) for the g.s.\ leads to a contradiction, compare
Fig.\ \ref{fig:angdistr} (a) and Fig.\ \ref{fig:tsys} (b); (iii) for the $0^+$ and $2^+$ states the orbital
momentum $l_y$ of $^{8}$He is coupled with the orbital momentum $l_x$ of two neutrons to total orbital
momentum $L=1$. Complete alignment of $L$ does not mean any specific alignment of the $^{8}$He orbital
momentum $l_y$. In contrast, for the $S=0$ configurations the $l_x=0$ dominance is expected and complete
alignment of the total orbital momentum $L$ is immediately transferred into a complete alignment of $^{8}$He
orbital momentum $l_y$. In this case the amplitudes for the angular distributions of $^{8}$He
in the selected $^{10}$He c.m.\ frame are obtained as a result of coherent summation of Legendre
polynomials $P_l^0(x)$. This provides the explanation for Eq.\ (\ref{eq:polinomial}).

Level ordering $0^+$, $1^-$, $2^+$ is inferred from the configuration choice
shown in Fig.\ \ref{fig:tsys} (a). Thus, the correlation data provide evidence for
anomalous level ordering in $^{10}$He. For the $1^-$ state the proposed correlation analysis
gives the energy and width around 5.5 and 2.5 MeV, respectively. For the $2^+$ state we can
establish only an energy range where the corresponding set of quantum numbers is important, see
Fig.\ \ref{fig:results} (d).

The interpretation of the data presented in this section is \textit{minimal required}. Indeed, the
Legendre polinomials with $l$ \textit{not less than} 1 are needed to describe the angular distribution in the
energy range $4<E_T<6$ MeV. Then, a visible asymmetry of the angular distribution presented in
Fig.\ \ref{fig:angdistr} (c) is a proof for the positive/negative parity state interference.
Taking into account the structural arguments of Fig.\ \ref{fig:tsys} we conclude that the $s/p$ interference
is a \textit{minimal required} set of components for this energy range. Analogous argumentation leads
to the $s/p/d$ assignment for the energy range $6<E_T<8$ MeV, see Fig.\ \ref{fig:angdistr} (e).
More complex interpretations would require much more conditions to be fulfilled strictly and
simultaneously.

%===============================================================================
%
\textit{Energy distributions in $^{10}$He.}
%
%===============================================================================
%
Additional qualitative support for the conclusion made about the population of different states at
$E_T<4$ MeV, $4<E_T<6$ MeV, and $6<E_T<8$ MeV can be found in the energy correlations occurring within these
ranges, see Fig.\ \ref{fig:angdistr} (b,d,f). The energy distribution parameter
$\varepsilon$ shows how the energy is shared between the $^{8}$He and ``dineutron'' subsystems.
This distribution is close to the phase volume for the ground state, see Fig.\
\ref{fig:angdistr} (b). However, it is qualitatively different for the excited states where the
``dineutron'' energy correlations ($\varepsilon \sim 0$) are enhanced. This effect is especially strong
for the energy range $4<E_T<6$ MeV where the $1^-$ state shows up. The observation of the expressed
``dineutron'' energy correlations is an additional argument supporting the $\Delta S=0$ transfer in our
experiment, see discussion of Fig.\ \ref{fig:tsys} above. We can expect the low-energy enhancement in
the $n$-$n$ channel in the case of attractive $n$-$n$ final state interaction available in the
$S=0$, $l_x=0$ configurations.

%===============================================================================
%
\textit{Theoretical calculations.}
%
%===============================================================================
%
The g.s.\ of $^{10}$He was extensively studied theoretically in Ref.\
\cite{Grigorenko:2008} focusing on a possible existence of a ``three-body virtual
state'' (an extremely low-energy peak with $[s^2_{1/2}]$ structure). Several
versions of calculations were provided depending on the scattering length in the
$^{8}$He-$n$ channel. The recent experimental results \cite{Golovkov:2007,Johansson:2010}
do not support the existence of a virtual state in $^{9}$He with a large negative scattering length.
So, considering the $^{10}$He g.s.\ we can stick to the predictions  based on the $^{8}$He-$n$
interactions with the $s$-wave scattering length around zero. Such calculations provide the g.s.\
of $^{10}$He with the dominant $[p^2_{1/2}]$ structure at about 2 MeV in a nice agreement
with the present experimental result, see Fig.\ \ref{fig:results} (e).

%-------------------------------------------------------------------------------
\begin{figure}
\centerline{\includegraphics[width=0.45\textwidth]{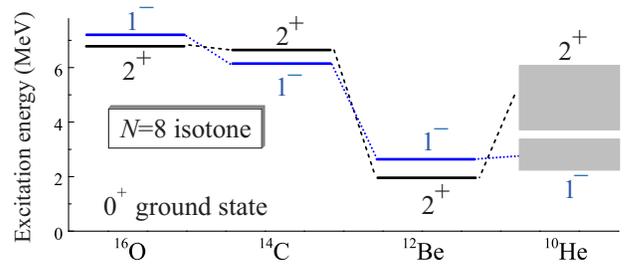}}
\caption{Low-lying $2^+$ and $1^-$ states for the $N=8$ isotone chain of $p$-shell nuclei.
Gray rectangles indicate the uncertainty of the $^{10}$He level positions in our
analysis.}
\label{fig:enn}
\end{figure}
%-------------------------------------------------------------------------------

According to calculations \cite{Grigorenko:2008} the results reported in
Refs.\ \cite{Korsheninnikov:1994,Johansson:2010} do not contradict the g.s.\ energy of
$^{10}$He obtained in the present work. In these works the $^{10}$He spectrum
was populated by the proton knockout from $^{11}$Li. It was demonstrated in
\cite{Grigorenko:2008} that the $^{10}$He g.s.\ observed at about $2.0-2.5$ MeV in
``conventional'' reactions [like the ($t$,$p$) transfer used in our work] should
have observable position $1.0-1.5$ MeV for the reactions with $^{11}$Li due to the strong
initial state effect. The observable g.s.\ peak position is shifted towards
lower energy because of the abnormal size of $^{11}$Li possessing one of the most
developed known neutron halos.

In this Letter we extend the calculations of Ref.\ \cite{Grigorenko:2008} to
the $1^-$ and $2^+$ excitations of $^{10}$He. The model predictions for the $^{10}$He spectrum
population give quite broad structures with very asymmetric shapes, see Fig.\ \ref{fig:results} (e).
The shaded areas show the ranges provided by theoretical calculations with a realistic parameter
variation. The calculation results are very stable for the g.s.\ energy but demonstrate increasing
uncertainty for the higher-lying excitation spectra. However, in all the calculations the lowest
excitation is $1^-$ providing additional support to the proposed experimental spin assignment.

%===============================================================================
%
\textit{Discussion.}
%
%===============================================================================
%
Observation of the $1^-$ configuration as the first excited state in $^{10}$He
is the most intriguing finding of this work. The existence of such low-lying
excitations is not something totally unexpected in the nearby exotic nuclei.
Such excitations in the form of \textit{soft dipole mode} are known for $^{6}$He
\cite{Aumann:2005}, $^{11}$Li \cite{Nakamura:2006} and there is evidence
that $1^-$ is the lowest excitation of $^{8}$He \cite{Golovkov:2009}.

The importance of the intruder configuration is evident in $^{11}$Be where the
existence of neutron halo is connected to the anomalous $1/2^+$ spin-parity of the ground
state. In the $^{12}$Be spectrum the breakdown of the $N=8$ shell closure is seen due
to the existence of the low-lying $1^-$ state. The importance of this phenomenon was
broadly discussed both from experimental \cite{Iwasaki:2000,Pain:2005} and
theoretical points of view \cite{Brown:2001,Gori:2005}. Our results provide novel information on
the evolution of low-lying level ordering of the $N=8$ isotones, see Fig.\ \ref{fig:enn}.
In $^{10}$He the $1^-$ state is found to be at the energy comparable
to that in $^{12}$Be, while the $2^+$ state is at the energy comparable to that
in the other members of the isotone chain.

%===============================================================================
%
\textit{Conclusions.}
%
%===============================================================================
%
The low-lying spectrum of $^{10}$He was studied in the transfer reaction $^3$H($^8$He,$p$)$^{10}$He.
The $0^+$ g.s.\ energy and width are found to be $2.1\pm0.2$ and $\sim 2$ MeV, respectively.
Owing to specific angular correlations for the first time the spin-parity assignment is made for
the low-lying states of $^{10}$He: the analysis of experimental data allowed to interpret the $^{10}$He
spectrum as a superposition of the $0^+$, $1^-$ ($E_T>4$ MeV) and $2^+$ ($E_T>6$ MeV) states. The established level
sequence shows that $^{10}$He is one more dripline nucleus demonstrating the shell structure breakdown.

%===============================================================================
%
\textit{Acknowledgments.}
%
%===============================================================================
%
The authors are grateful to Profs. B. Jonson, M.V. Zhukov, S.N. Ershov and I.G. Mukha for useful
discussions. This work was supported by the Russian RFBR 11-02-00657-a grant. L.V.G., S.A.K., A.V.G., and
I.A.E.\ are supported by FAIR-Russia Research Center grant. L.V.G. acknowledges the
support by HIC for FAIR research grant, and Russian Ministry of Industry and Science grant
NSh-7235.2010.2.

%###############################################################################

%###############################################################################

\end{document}